\documentclass[fleqn,10pt,onecolumn]{SelfArx} 
\usepackage{graphicx}
\usepackage{hyperref}

\definecolor{color1}{RGB}{0,0,90}  
\definecolor{color2}{RGB}{0,20,20} 

\usepackage{hyperref} 
\hypersetup{hidelinks,colorlinks,breaklinks=true,urlcolor=color2,citecolor=color1,linkcolor=color1,bookmarksopen=false,pdftitle={Title},pdfauthor={Author}}

\JournalInfo{\mbox{\;}} 
\Archive{} 

\PaperTitle{ConGUSTo: (HT)Condor Graphical Unified Supervising Tool} 

\Authors{
    Antonio Dorta\textsuperscript{1,2}*, 
    Nicola Caon\textsuperscript{1,2}, 
    Jorge Andr{\'e}s P{\'e}rez Prieto\textsuperscript{1,2}
}

\affiliation{\textsuperscript{1}\textit{Instituto de Astrof{\'\i}sica de Canarias, E-38205 La Laguna, Tenerife, Spain}} 
\affiliation{\textsuperscript{2}\textit{Universidad de La Laguna, Dpto. Astrof{\'\i}sica, E-38206 La Laguna, Tenerife, Spain}} 
\affiliation{*\textbf{Corresponding author}: adorta@iac.es} 

\newcommand{\keywordname}{Keywords} 

\Abstract{ 
HTCondor is a distributed job scheduler developed by the
University of Wisconsin-Madison, which allows users to run their applications
in other users' machines when they are not being used, thus providing a
considerably increase in the overall computational power and a more efficient
use of the computing resources. Our institution has been successfully using
HTCondor for more than ten years, and HTCondor is nowadays the most used
Supercomputing resource we have. Although HTCondor provides a wide range of
tools and options for its management and administration, there are currently
no tools that can show detailed usage information and statistics in a clear,
easy to interpret, interactive set of graphics displays. For this reason, we
have developed ConGUSTo, a web-based tool that allows to collect HTCondor usage
and statistics data in an easy way, and present them using a variety of
tabular and graphics charts. 
}

\begin{document}

\def\UrlFont{\sl}
\onecolumn

\flushbottom 

\maketitle 

\tableofcontents 

\thispagestyle{empty} 


\section{Introduction} 


The Instituto de Astrof{\'\i}sica de Canarias (IAC) is the largest Spanish
astrophysical research centre. The IAC manages two observatories located over
2000 meters ASL with more than 25 telescopes on the whole, including the
current largest optical/infrared telescope in the world (GTC), all equipped
with state-of-the-art instrumentation. These telescopes generate large
amounts of raw data which must be processed and reduced before they can be
analyzed. The data processing is usually done with both general-purpose and
specialized software packages that need a lot of computational power. Most of
this data reduction work is carried out by applying the same tasks over a
large number of independent inputs, so it is much more suitable for
distributing computing than for parallel supercomputing. In fact, there are
several research projects at the IAC  which are viable only thanks to the use
of these kind of resources, since they need several thousands of CPU hours to
be completed.\footnote{As an example, one of the IAC research group is using
holography techniques to create seismic maps of the solar interior, from
observed velocity fields obtained by GONG (Global Oscillation Network Group).
Maps are produced for about 500 depths, for observations spanning several
months in order to study temporal variations. Each map requires about 10
minutes of computing time:  HTCondor is the ideal resource for such kind of
work, as it involves applying the same process to thousands of independent
inputs.} So HTCondor, a software tool which will be briefly described in the
following sections, plays a fundamental role at the IAC, with over 200 Linux
desktops and 700 computing slots available to researchers when they are not
being used by their owners. In the first semester of 2014, our staff executed
jobs in HTCondor for a total of about 1,300,000 computing hours, that is the
equivalent of one and a half century of sequential computing time in just six
months.

As HTCondor system administrators, our interest in monitoring and supervising
how HTCondor works is twofold. First, we must assure that HTCondor does not
interfere with the regular use of the desktop PC by the user who is logged
in. Both an imperfect knowledge of what HTCondor does (and what doesn't), and
some past failures in detecting user's activity and evicting HTCondor jobs (a
problem caused by an Operating System upgrade and fixed shortly after it
showed up) have made the users suspicious of HTCondor, which they have
started to blame whenever their computer becomes slow, unresponsive or
freezes at times (HTCondor has thus become the favorite scapegoat). Second,
we must respond to the energy saving measures that are being studied and
implemented by the IAC in order to reduce electricity bills. One of the
proposals was to turn off all Linux desktops at night and on weekends, and we
need to show that most of such desktops are indeed used by HTCondor outside
regular work hours, and we cannot afford turning them off without seriously
compromising those research activities that heavily rely on HTCondor.

The standard HTCondor installation comes with several tools and options for
monitoring how each job was executed and to produce a log that can be
interpreted by some HTCondor commands and external applications. It also
generates some general statistics about the pool and the allocated resources.
However, we have found out that there is no easy way to track the detailed
usage of each machine by HTCondor, which is a key factor for our purposes.

In particular, we are interested mainly in the following points:

\begin{itemize}

\item To obtain detailed information about what jobs have been executed in
each of our machines, at any time, with all the associated information such
as owners, states, etc., as well as a comprehensive summary encompassing all
machines. Such information must be easily accessed, preferably shown online
in a graphic way easy to interpret and understand, thereby allowing users to
check the status of their machines and contact us when experiencing issues
that could be related to HTCondor.  

\item To collect further information about when each machine is executing jobs,
with the aim of implementing some power saving measures like automatically
suspending or hibernating idle machines, and letting HTCondor wake them upon
jobs allocation.

\item To have a real-time graphical "panoramic view" of the status of all our
machines. We do not wish to merely create a graphical representation of the
\textsl{condor\_status} command output, but we want also to include
additional information that is not generated by that command and could be
very useful to detect problems: the last time that a specific machine
executed a job, if it has some specific time restriction (for instance
HTCondor jobs are allowed to run at night and on weekends only), if there is
enough space in the scratch disk, and many other parameters that allow us
to do an in-depth assessment of individual machines.

\item To avoid having to install additional software on each machine that
belong to HTCondor pool.

\end{itemize}

With these goals in mind, we have developed ConGUSTo, a graphical online
supervising and statistics system for HTCondor that we believe could be
interesting for other HTCondor users and administrators. This paper presents
ConGUSTo, and is organized as follows: Section 2 gives a brief description of
HTCondor and the current log options that it offers, while in Section 3 we
list the main ConGUSTo features. In Section 4 we present how ConGUSTo has
been implemented, and in Section 5 several illustrative examples are shown.
Finally, Section 6 gathers our conclusions and proposes future lines of
development.

\section{HTCondor and its current log options}

This section does not try to be a rigorous or exhaustive description of
HTCondor and all its log possibilities; any user interested in that can
consult the extensive documentation available at
[\url{http://research.cs.wisc.edu/htcondor/}]. Our intent here is to offer a
very short account of how HTCondor works, and explain briefly those common
commands and options available that can be used to obtain information about
jobs and machines' status.

\subsection{What is HTCondor?}

HTCondor is an open source software that gives support to High Throughput
Computing (HTC) [\url{http://research.cs.wisc.edu/htcondor/htc.html}]. It was
developed by the Center for High Throughput Computing at University of
Wisconsin-Madison[\url{http://chtc.cs.wisc.edu/}]. In a nutshell, according
to its developers, it is "a specialized workload management system for
compute-intensive jobs". Basically, HTCondor manages a pool of machines and a
queue of jobs: whenever a machine is idle (not being used by their owners and
with a low load) HTCondor will match a suitable job to run on it.

HTCondor will carry out all the operations needed to correctly match
idle machines with queued jobs according to specifications coming from both,
execute those jobs and place the outputs where the user has specified,
successfully dealing with a wide and complex set of circumstances.

Thus, using the HTCondor software, a group of off-the-shelf PCs connected by
a local network can be transformed into a powerful Supercomputer where users
can simultaneously run several instances of their programs over a variety of
data, reducing considerably the time needed to obtain the results (up to a
factor of several hundreds, depending on the application, the data and the
number of available machines). Besides, HTCondor not only provides a
substantial boost to computational capacity, but also enables a better and
much more efficient use of the available computing facilities, minimizing the
waste of energy and resources. HTCondor is used in many companies and
institutions, among them the NASA Advanced Supercomputing Division, the
European Organization for Nuclear Research (CERN) 
[\url{http://linuxsoft.cern.ch/cern/mrg/slc5X/x86\_64/RPMS/repoview/condor.html}],
etc. A more detailed, technical introduction of HTCondor can be found at
[\url{http://research.cs.wisc.edu/htcondor/description.html}].

\subsection{HTCondor logging and statistics options}

HTCondor has several options in its configuration files to manage logfiles,
which are generated and updated by the different daemons that HTCondor runs.
By default, each computing machine keeps its own logs, with HTCondor providing
a command that allows administrators to consult log files located on those
machines. As usual, administrators can set the log level to decide what
information will be stored in the logfiles, specify the place where they will
be created, impose a size limit, etc.

Users running their jobs with HTCondor can include a command in their submit
files in order to generate specific log files for each execution. Once the
jobs are done and the logs generated, they can be the used to get detailed
information about the execution. HTCondor provides a set of commands for that
purpose, such as \textsl{condor\_userlog}, \textsl{condor\_history}, etc. If
users are interested in getting a graphical visualization of such data, they 
can use the java application logview or some third-party
applications like the HTCondor Log Analyzer
[\url{http://condorlog.cse.nd.edu/}].

Finally, if enabled, HTCondor is able to generate some general statistics,
which are stored in plain HTML format and can be displayed graphically by
means of a java applet. There are two kinds of information: the first is a
graph and tabular data for the Pool User (Job) Statistics (total allocation
time, average and peaks of running and idle jobs), the second the Pool
Resource (Machine) Statistics (average and peak number of machines according
to their states). Information is given for the current day, week, month, or
any month in the last year.

The latest stable versions of HTCondor that have been recently released
(8.2.x) claim to have an enhanced monitoring system
integrated with the third-part software Ganglia [\url{http://ganglia.info/}].
We have tested the live demo and found it much more powerful that any of the
previous log and statistics systems available in HTCondor. However, it does
not fulfill our requirements, described above, as it seems to be too focused
on showing those physical parameters contained in the ClassAds, and it does
not appear to be compatible with previous versions of HTCondor. There are
other third-part software alternatives, like CycleServer
[\url{http://www.cyclecomputing.com/products-solutions/cycleserver}] or Cumin
[\url{https://fedorahosted.org/grid/wiki/CuminForUsers}], but although they
can be powerful, they also need a more complex management and/or the
installation of software on each machine of the HTCondor pool.

\section{What ConGUSTo offers}

As we have just explained, the present HTCondor logging system is useful for
debugging or check details about executions; however, trying to know, for
instance, if a machine executed some job a few days (or months) ago and what
and whose job it was could be really difficult, as it may require to dive
into huge log files all with a somewhat complicated format, or it may be even
impossible if the log data are too old, since logfiles are periodically
deleted to save space.

Motivated by these limitations, we have developed ConGUSTo, which offers many
features not available in the current HTCondor implementations. The most
important are:

\begin{itemize}

\item Graphical view of all jobs executed in a specific machine, with more
detailed info about each of them accessible online (using any web browser).
For each machine in the pool we can show both summarized and detailed info
about all the jobs that HTCondor has executed on it.

\item Friendly and intuitive way to display data. Besides tables with
numerical data, most information is also presented by means of interactive
graphs.

\item Data can be presented in periods of one day, one week or one month,
allowing users and administrators as well to choose the starting day of that
period.

\item Panoramic view of all machines and slots to see the current HTCondor
status in a glance.

\item Panoramic view is highly configurable with a wide range of options and
filters.

\item Easy installation: ConGUSTo only needs to be installed on one machine
(typically the web server). No data base is required: the data, stored in
plain text with a format similar to CSV, are generated by just one of the
machines in the pool (in our case it is the HTCondor master node, which is on
24/7) using a simple bash script executed by crontab. Thus there is no need for
HTCondor to be installed on the same machine where ConGUSTo is located.
\newline

\item Fully compatible with previous versions of HTCondor.

\item It shows data that at the present moment Ganglia cannot supply, such as
information per slot, time restrictions, last time a machine executed a HTCondor
job, and much more.

\item Adding new pieces of information to show is easy.

\end{itemize}

\section{Implementation of ConGUSTo}

As already mentioned, our objective is to generate a more friendly log system
based on the information that we can obtain from HTCondor. We want this
system to be able to show, in a intuitive and interactive way, what each
machine was doing at any time, if it was executing a job (and then which job
and by whom), whether it was idle, or used by someone else. These data must
be kept updated and available online in order for both HTCondor
administrators and users to be able to visualize them.

\subsection{How ConGUSTo retrieves the data}

The jobs data we want to show in principle could be extracted from the log
files generated by HTCondor. For instance, we could parse the logs generated
by the daemon \textsl{STARTD} on each machine that executes jobs, in order to
track what jobs have been executed in each slot. However, we decided to
discard this option for a number of reasons: such files generally have a
different location depending on each machine; they are usually quite big and
it may take a long while to parse them; their specific format in plain text
is quite complex and not suitable for this purpose; while they contain many
useful data, some basic, interesting info is missing, for instance the status
of the job. User's logs obtained when executing are not useful either, since
they depend on the users' commands written in the submit files, and even if
created, they will be located wherever the user has specified. Finally, the
statistics generated by HTCondor are useful for knowing the global status of
the jobs queue or the machines' pool at a given time, but cannot track what
each machine has been doing.

Thus, we have implemented our own system to obtain the required data. Instead
of reading the data off the log files, we obtain the information using
standard HTCondor shell commands and then we process their output. We only need to
execute the command once to get the information from all machines, avoiding
parsing hundreds of huge log files, one for each machine. 

In this way we obtain all the information we need, restricting it to those
jobs that are running or suspended, the ones we are interested in. The output
from this command is then processed by a shell script using simple shell
commands such as \textsl{grep}, \textsl{sed}, \textsl{awk}, \textsl{tr},
\textsl{cut}, etc, and finally stored in a CSV-like format. Everything is
managed by a bash script periodically executed by \textsl{crontab}. This
script has fewer than 20 lines of code, which reflects the efficiency of our
approach.

ConGUSTo has been designed in such a way that data displaying is totally
independent from data generation. This perspective has the advantage of not
requiring the web server to run HTCondor itself (we think that web servers
should be doing their work and not running unrelated software packages), and
we can use any other machine where HTCondor is running to obtain all data
(typically the HTCondor collector). We evaluated the option of storing the
collected data in a database, but to make it simple and to avoid the need for
a database system, we chose to store the data in plain text files keeping the
format as it is generated (CSV-like). So to let it work, we only need to make
sure that data are stored in a location accessible by the web server. For
example, files can be created in a shared directory accessible from the web
server via NFS, or transferred by FTP, etc. The file organization is also
simple: there is a root directory and then subdirectories for years, months
and days, in a tree structure where the nodes are files of a concrete date,
one file per each machine. This organization pattern based on dates and
machines makes it easy to find the data for a specific date or range, and
also makes the installation of ConGUSTo in other systems much simpler.

On the other hand, there is also some information that does not need to be
stored, for instance the current state of machines in the pool. This kind of
data is obtained in real-time and since it is directly displayed, there is no
need to save it. (Those data are obtained by the use of the
\textsl{condor\_status} command, forcing it to produce its output with a
predetermined format.)


All the commands and applications that ConGUSTo uses to get and store the data
are (or should be) available in any Linux machine. We also expect that
implementing the same tools in Windows should be relatively straightforward.

\subsection{How ConGUSTo displays the data}

ConGUSTo has been developed from scratch. To generate the web pages that display
the information we have chosen PHP [\url{http://php.net/}]. PHP is a web
programming language widely popular, available on virtually all Operating
Systems, and used by such big enterprises or web projects as Wikipedia,
Yahoo, (some parts of) Facebook, Wordpress, etc.
[\url{http://en.wikipedia.org/wiki/PHP}]. As it can be run in Linux and
Windows Server (and Mac OS X as well), it helps make our implementation very
portable.

All the calculations needed to process and display the data are done by PHP
scripts, while graphs are generated using the API of Google Charts
[\url{https://developers.google.com/chart/}], also widely used and
customizable. This allows us to show rich, interactive graphs where users can
consult specific items. ConGUSTo translates the information stored in data files
into a representation that Google Charts can understand in order to draw the
graphs. We have customized this translation process in such a way that adding
new data and graphs is easy: most times it is sufficient to specify the new
graph position and type to automatically generate it. This proved
to be especially useful in the code that shows a panoramic view of the
current status of all machines, where dozens of options have been implemented
to show or hide different kinds of data for every machine and/or slot, filter
them according a wide set of attributes, or select among a large number of
predefined graphs. In particular, user can choose to show or hide three main
groups of information (machines, queue, graphs about attributes), while each
of these groups can be independently customized. For machines, a lot of
different data are available, like name, number of slots, free disk space
(total and per slot), memory (total and per slot), Operating System, average
load (both total and HTCondor), etc. Besides this general information, our
tool also stores some specific data, one of the strong points of our
implementation, since it is not commonly offered by other tools: time
restrictions on jobs execution, last time a job was executed on each machine,
etc.

To finish, we want to mention other third-party tools and snippets we have
used in our development, such as \textsl{SCW} (Simple Calendar Widget) to present the
calendar [\url{http://www.tarrget.info/calendar/scw.htm}], \textsl{Tooltip}
[\url{http://jqueryui.com/tooltip/}] to display extra information according
to the context, \textsl{Chosen} to allow searches on the lists
[\url{https://github.com/harvesthq/chosen}] and several CSS generators.

\section{Some examples}

We present in this section several screenshots of ConGUSTo to illustrate its
functionalities and main features. 

\begin{figure}[ht]\centering
\includegraphics[width=\linewidth]{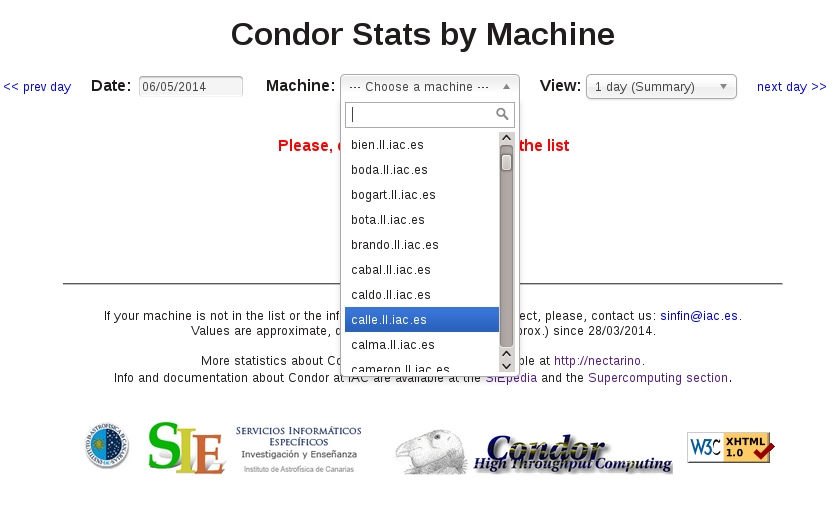}
\caption{Users can select a specific machine for which to display usage 
statistics.}
\label{fig:choose_machine}
\end{figure}

The first contact you have with our tool is a screen where you can choose
which date, machine and view you want to see, as shown in
Fig~\ref{fig:choose_machine}. All the options are displayed in a friendly
style: users can choose the date pressing on a navigable calendar, while
machines can be selected from a list that includes a search box, so it is
even easy in large systems with hundreds of machines. Several views are
available and they will be explained next.

\subsection{Daily jobs execution on a machine (summary)}

\begin{figure}[ht]\centering
\includegraphics[width=\linewidth]{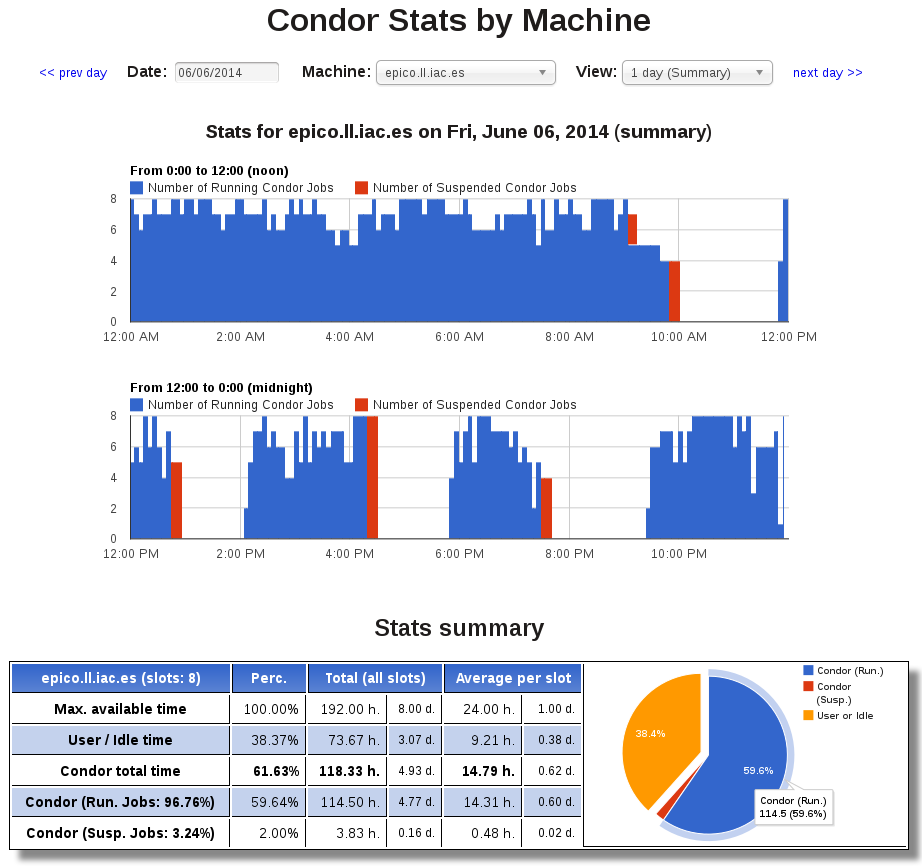}
\caption{Usage statistics for machine "epico" for a specific day. 
The vertical axis represents the number of CPU cores (8 in this case).}
\label{fig:condor_summary2}
\end{figure}

This first view will present summarized information about the selected day,
as is shown in Fig.~\ref{fig:condor_summary2}. Two graphs are displayed with
a timeline listing the number of jobs executed at any moment (upper graph
displays the AM hours and lower one the PM). The number of jobs that are or
were running are shown in blue, while those in red are the suspended ones.
This graph gives us in a glance an overview of what HTCondor has been doing
on that machine, when it has been executing jobs and when some activity has
been detected and jobs were suspended.

After the graphs, users can find a table that gathers quantitative data about
the total time spent. The first row shows the maximum theoretical time
limit;  the second row displays the total amount of time the machine has been
used by its owner or was idle (total time is printed alongside the average
per slot and percentages), while the third row collects the total time
consumed by HTCondor,  broken down into time used when jobs are running and
when are suspended (fourth and fifth rows, respectively).

\subsection{Daily jobs execution in a machine (details)}

\begin{figure}[ht]\centering
\includegraphics[width=\linewidth]{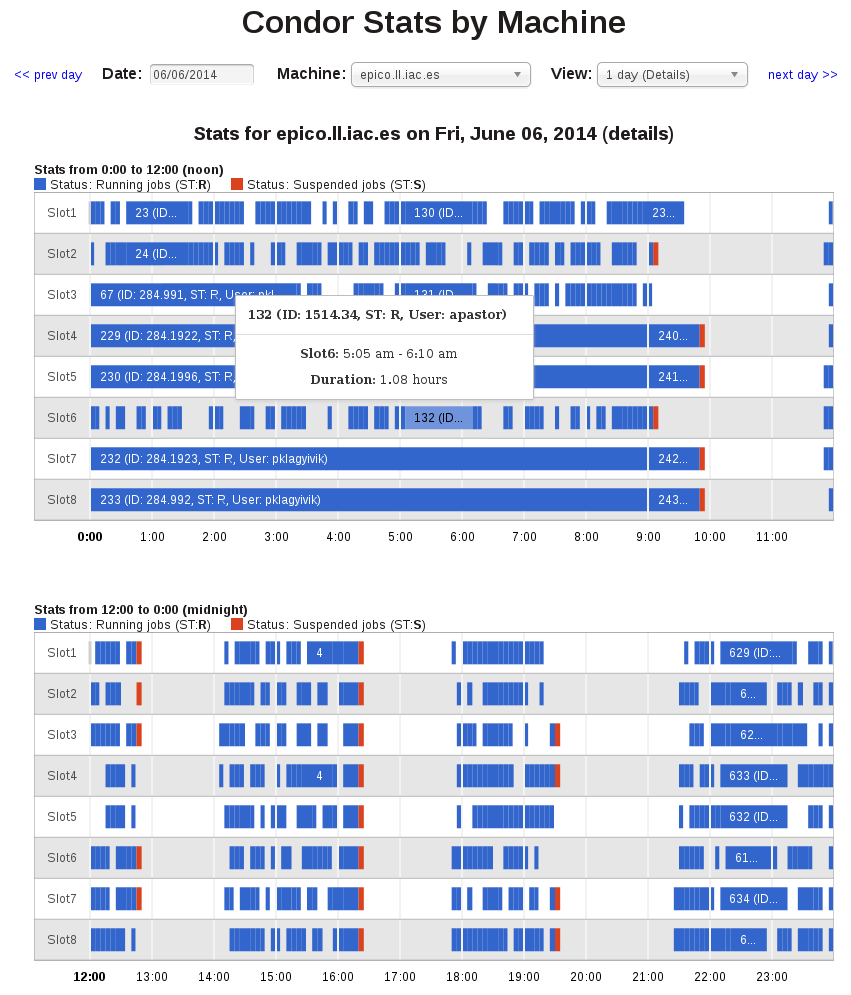}
\caption{Graphical representation of all jobs that were executed 
on "epico". By positioning the cursor on a specific job, a tooltip is 
produced with detailed information.}
\label{fig:condor_status_details2}
\end{figure}

If users want further information about the execution on an specific day,
they can choose the detailed view as shown in
Fig.~\ref{fig:condor_status_details2}. This view is similar to the summary, but
graphs show detailed information about each job that was executed on each
slot of the selected machine. As in the summary, for a clearer view the
timeline is split between AM and PM hours, and jobs are shown in blue when
they are running and in red when suspended. Among the information displayed
is the ID of the job, the current status, its user, initial and ending time,
and the total duration. If the job is long enough and there is space, part of
that information will be printed on the job. If not, all information will be
shown in a small pop-up window when you place the pointer over the job.
Thanks to this view one can get such data as what jobs and users were
executing on an machine, and use this information to gather further data in
case of any issue.

\subsection{Weekly and monthly jobs execution in a machine}

\begin{figure}[ht]\centering
\includegraphics[width=\linewidth]{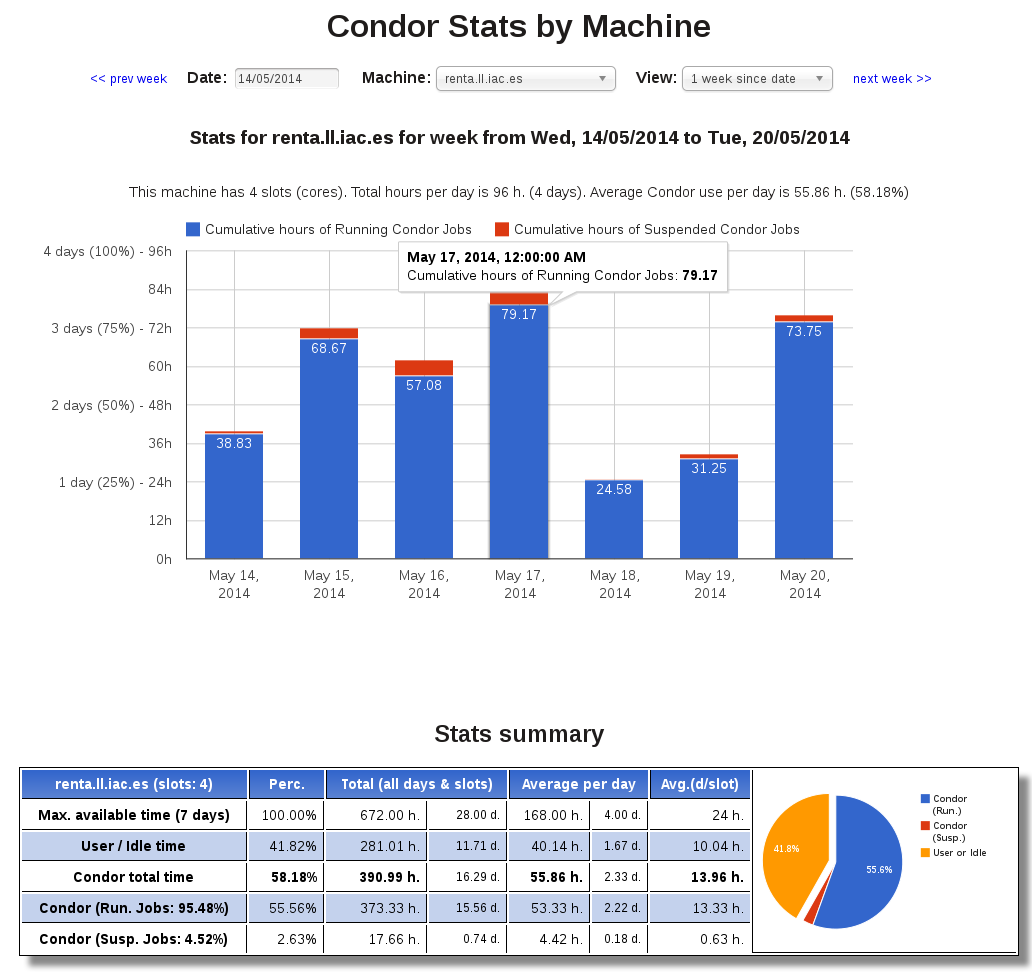}
\caption{Condor usage over a one-week period for machine "renta".  
Blue color represents the time consumed by running jobs, while red color is 
used for the suspended ones. Summary table shows total values, together with 
additional data to make comparisons easier, like the average data per day 
and per day/slot.}
\label{fig:1week}
\end{figure}

Besides the summarized and detailed information for specific days, our tool
also offers an overview of HTCondor usage for one-week and one-month periods.
Users can choose the starting date of those periods and then display the
cumulative hours and percentage of time consumed by HTCondor. Fig.~\ref{fig:1week}
illustrates the one-week-period in both graphical and tabular way.

\begin{figure}[ht]\centering
\includegraphics[width=\linewidth]{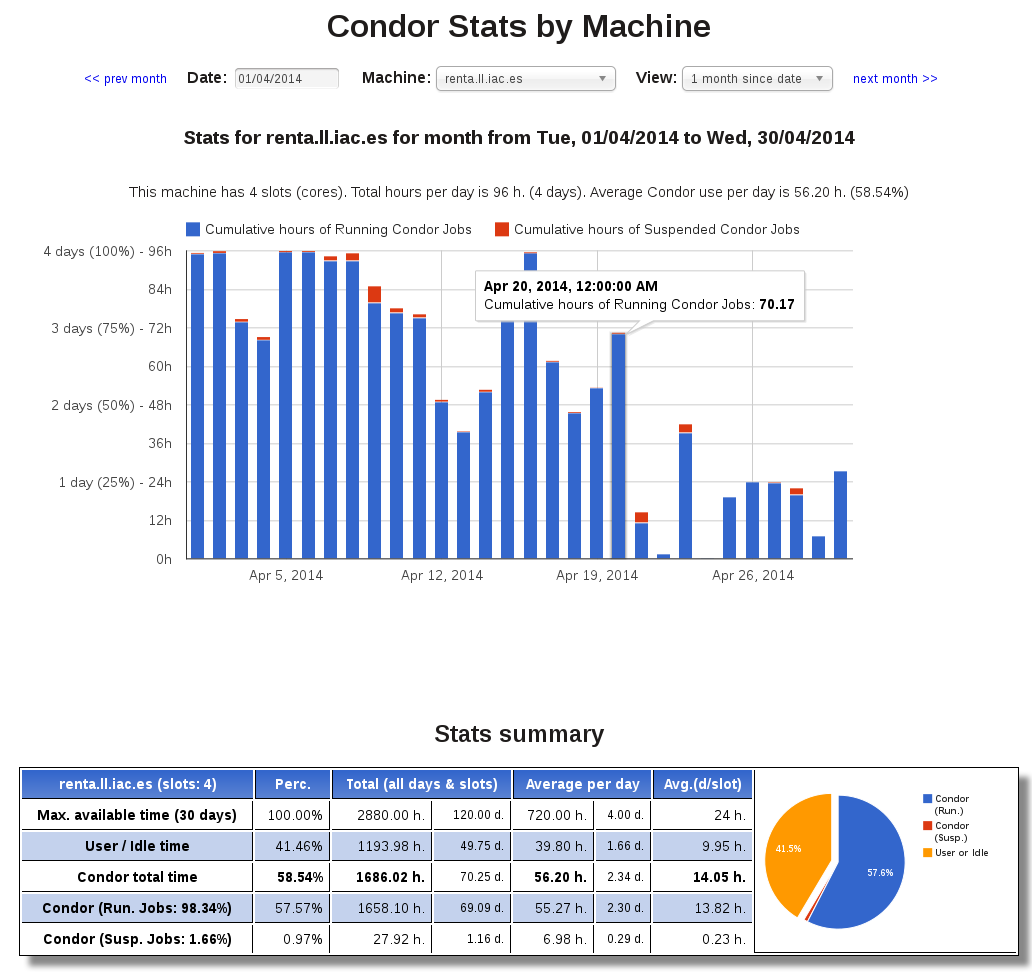}
\caption{Condor usage data over a one-month period.}
\label{fig:1month}
\end{figure}

The month view is similar to the week one, as it is shown in
Fig.~\ref{fig:1month}. It offers a wider view of the performance of that
machine,  so it is easy to identify the days of highest or lowest usage, and
then inspect summarized and/or detailed information about them with other
views.

\subsection{HTCondor machine panoramic view}

\begin{figure}[ht]\centering
\includegraphics[width=\linewidth]{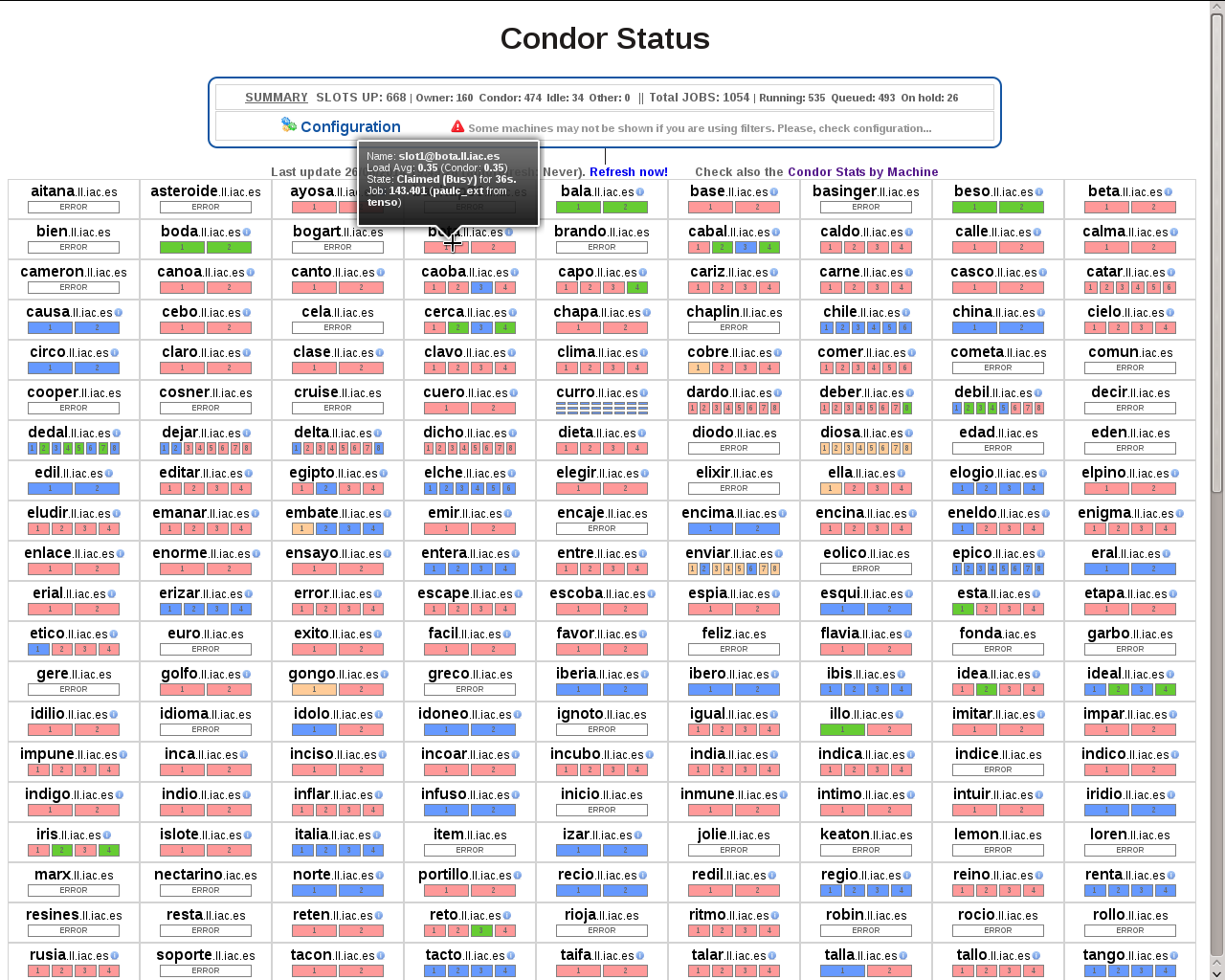}
\caption{The default panel shows the status of each machine in the HTCondor 
pool.}
\label{fig:condor_status_pano}
\end{figure}

The previous views allowed us to know the past and current status of a
specific machine. Now we present a powerful, panoramic view that lets us know
in a glance the global status, at the current time, of all machines in the
HTCondor pool. This view can be seen as a complement of the
\textsl{condor\_status}  command, where the data are graphically displayed
instead of being printed in a large text list, and with many more featured
added. Fig.~\ref{fig:condor_status_pano} shows the default, basic view.

\begin{figure}[ht]\centering
\includegraphics[width=\linewidth]{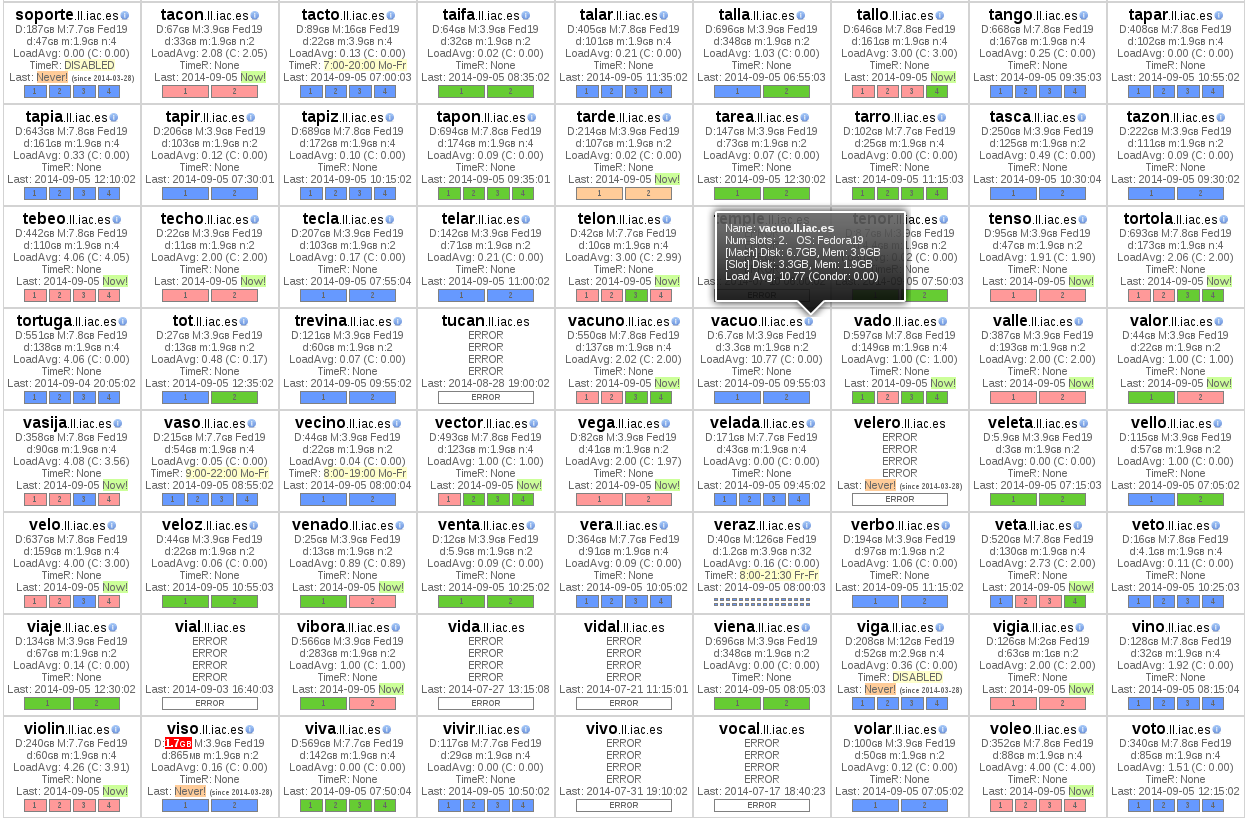}
\caption{Same as previous Figure, but with additional data for each machine.}
\label{fig:condor_status_detail3}
\end{figure}

Machines are represented with boxes and each slot is displayed with a
different color according to its state (blue when they are used by their
owners, red when running HTCondor jobs, green when idle, etc.), preceded by
a summary of the pool general status and the link to the configuration
options. By default only the name of the machine and the state of slots are
displayed, since our aim is to keep the visualization as clean as possible.
Therefore, detailed information is not presented by default, but our tool is
able to gather much more information that users can consult by just placing
the pointer over one specific machine or slot. Then extra info is displayed
in a contextual window: free disk space, RAM, number of slots, Operating
System, total average load and HTCondor average load, time restrictions, last
time of HTCondor execution, and many other indicators. However, if users
wants to have all these data in the main window, they can use the wide set of
provided configuration options to exactly select the specific items to be
displayed, as well as enable or disable the contextual windows. For
instance, Fig.~\ref{fig:condor_status_detail3} shows a more detailed view of the
pool status.

\begin{figure}[ht]\centering
\includegraphics[width=\linewidth]{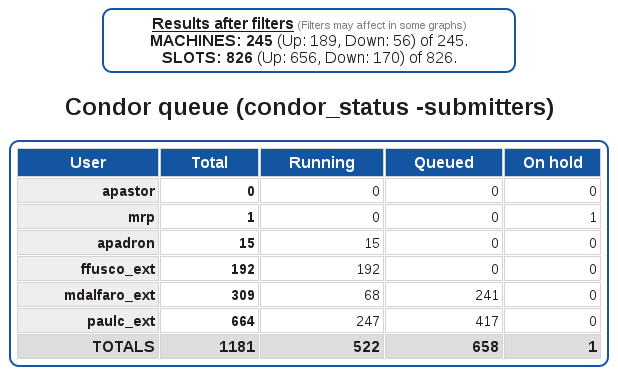}
\caption{Summary of the current status of HTCondor usage, broken down by status and users.}
\label{fig:condor_status_queue_info}
\end{figure}

The above data are complemented with information about the number of machines
and slots shown and the HTCondor queue status that is shown right after.
Fig.~\ref{fig:condor_status_queue_info} shows the summary of the queue, similar
to the output of the \textsl{condor\_status -submitters} command. It consists
of a table listing the total number of jobs, further broken down by their
state (running, queued or held) and by their users.

\begin{figure}[ht]\centering
\includegraphics[width=\linewidth]{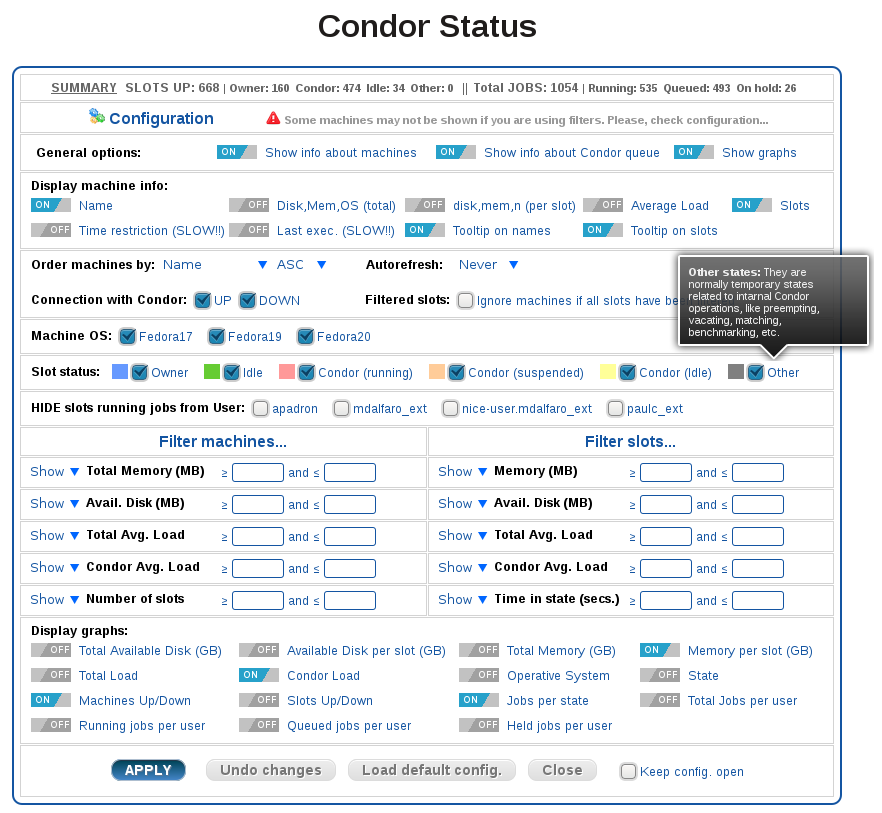}
\caption{Configuration panel with the list of all applicable options and 
filters to customize the panoramic view.}
\label{fig:condor_status_config}
\end{figure}

The real power of this tool lies in its configuration capabilities and in the
many available filters. Fig.~\ref{fig:condor_status_config} shows the
configuration window, which begins with general options to enable or disable
the three main groups: info about machines, info about HTCondor queue, and
graphs to represent the data. Users can configure which information and how
they want to display for each machine and/or slot, such as hardware
parameters (free disk, RAM, etc.), software (OS, average load, etc.), and
HTCondor specific data (time restrictions, last time of execution, etc.).
Further options let users to order the machines according to several
criteria, display alerts if the free disk space is lower than a specified
limit, or enable the automatic update of the information in preconfigured
time periods. 

Next, blocks are used to apply filters on the machines, allowing users to
have a more refined control of the HTCondor pool. In this way users can
decide which machines and/or slots they want to display according to their
communication status (up or down), which Operating System and version is
installed, the state of the slots, what user is running what job on them,
etc. Right after, there is a battery of filters to show or hide machines or
slots according to attributes like memory, free disk space, system average
load, HTCondor average load and number of slots (machines) or time spent in
the current status (slots). All these filters accept a range of values.

\begin{figure}[ht]\centering
\includegraphics[width=0.46\linewidth]{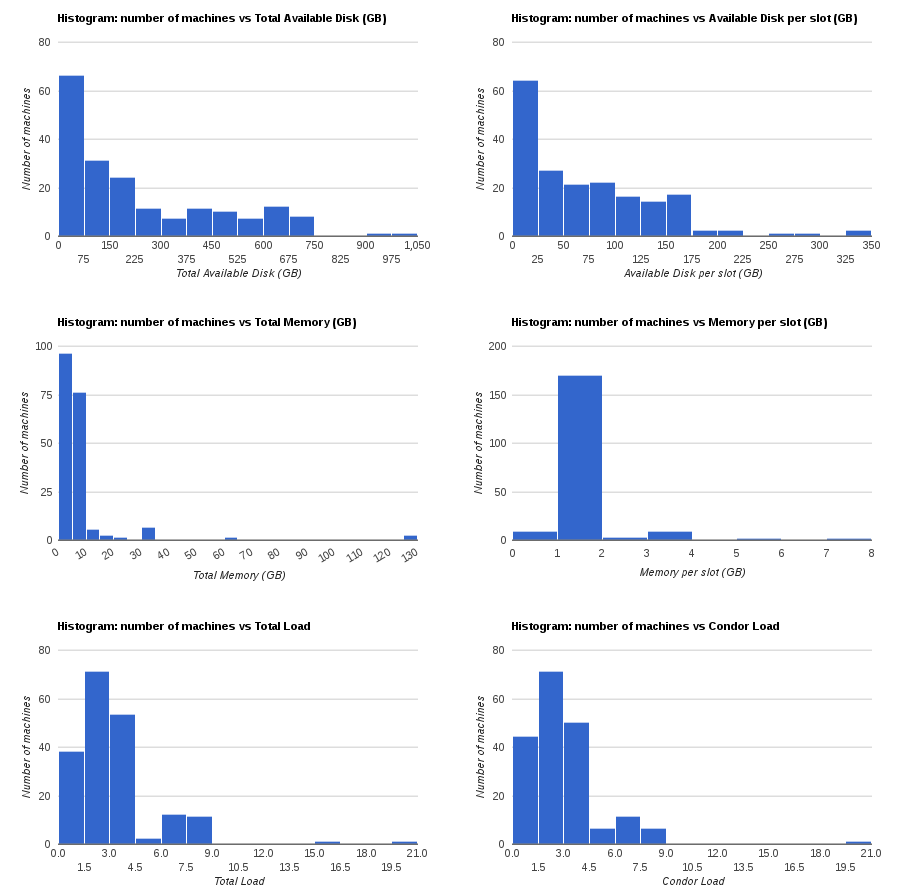}
\includegraphics[width=0.46\linewidth]{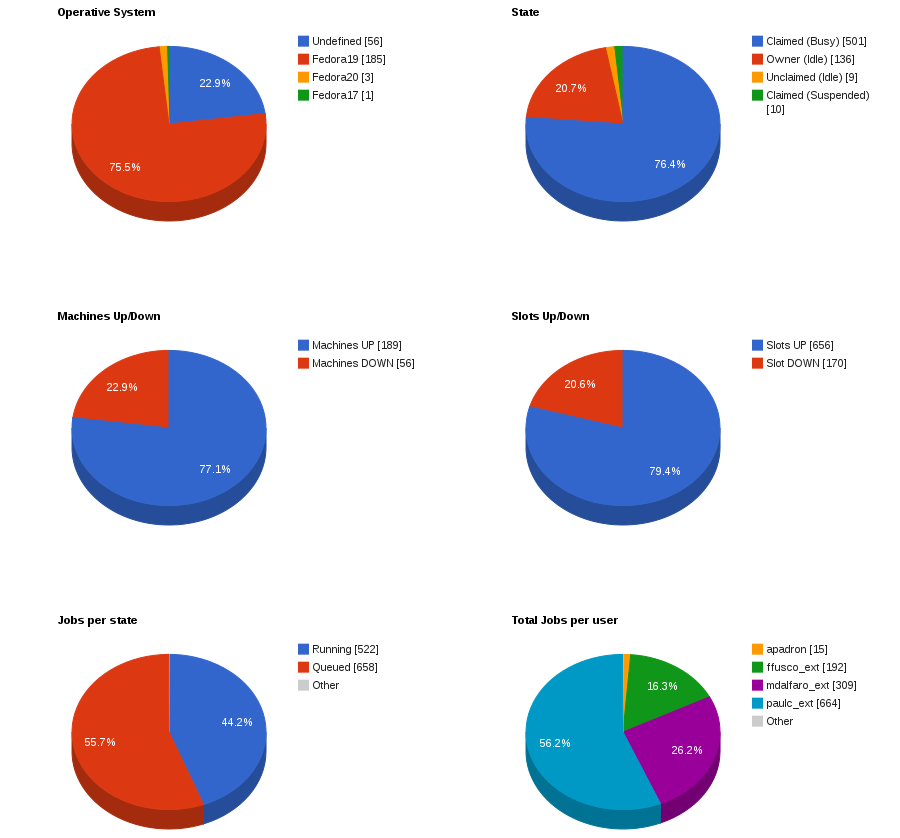}
\includegraphics[width=0.5\linewidth]{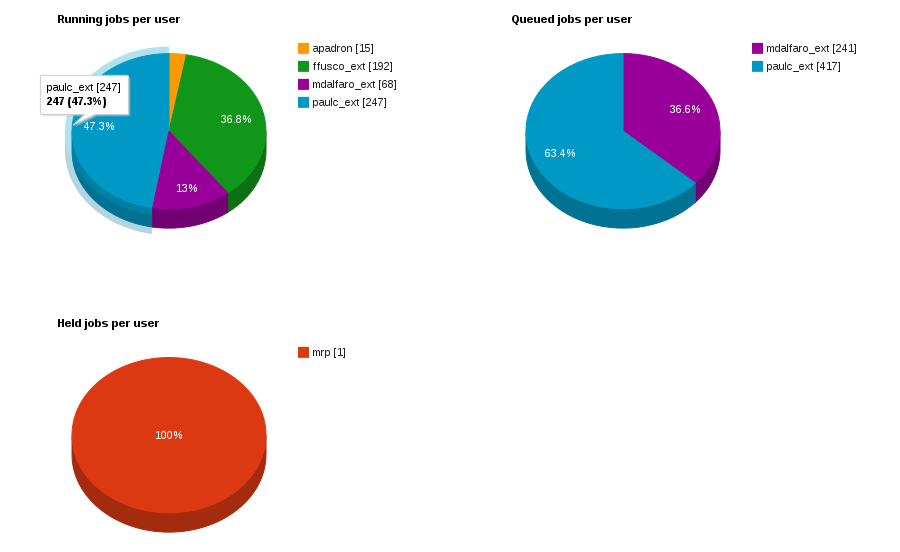}
\caption{Charts summarizing the current HTCondor usage.}
\label{fig:condor_status_graph}
\end{figure}

Finally, users can configure what data they want to show in charts. There is
a set of 15 predefined graphs to see the pool status according to a number of
parameters: amount of free disk or RAM, load, job state and owners.
Fig.~\ref{fig:condor_status_graph} shows some examples.

We want to point out that all these options are not only useful for HTCondor
users, but also for system administrators. With ConGUSTo, they are able to have
a graphical view of the machines status and get such valuable information as
what machines are up or down, average load of each machine, free disk and
total memory, Operating System, number of slots, etc. They can use also the
available filters to select  subgroup of machines, while the automatic
refresh feature can convert this application in a real-time monitoring tool.

\section{Conclusions and future work}

For many years HTCondor administrators have lacked a powerful tool that
provide a quick and clear overview of where and when HTCondor was executing
the jobs, and what was happening in each specific machine of the pool. The
new version in the 8.2.x series, released recently, is a good step in this
direction, as it includes new features that are able to generate data that
can be then represented with Ganglia, a third-party software. This feature is
expected to improve in subsequent releases, but right now it does not seem to
be compatible with older versions of HTCondor.

We have thus developed and implemented  ConGUSTo, an advanced logging and
monitoring system for HTCondor that provides an online visualization of the
status of all the machines that belongs to the HTCondor pool. Our
implementation is compatible not only with the most recent HTCondor release,
but also with older versions. Its main goal is to allow users and
administrators to know at any moment, in real time, what the status of any
machine is and what jobs they have executed or are executing. All this
information is available online in a friendly and easy to interact with
format, with no need to dig into huge log files anymore.

The current ConGUSTo implementation has room for a number of improvements. We
plan to explore the following lines: 

\begin{itemize}

\item Study other ways to store in a more compact format the gathered data. 

\item Test our implementation on other HTCondor systems with different
versions.

\item Identify what options and configurations are the most used by our
users. 

\item Receive feedback and comments from our users, so as to find new
features that could be implemented. 

\item Assess in detail the capabilities  provided in the latest HTCondor
release and see how they can be incorporated into ConGUSTo to enhance and expand
it.

\end{itemize}

\section*{Software availability}
\addcontentsline{toc}{section}{Software availability} 

While at the moment ConGUSTo is not publicly available, interested users can
contact us via email to get a copy. We will be more than happy if you wish to
share your impressions and feedback with us.

ConGUSTo is distributed in a \textsl{.tar.gz} file. Detailed installation
instructions will be provided to interested parties, however the basic steps
are: 

\begin{enumerate}

\item Decompress the tar file in a directory of your web server where you can
run PHP script files. 

\item Build the file containing the list of machines (this is used to list
all machines belonging to the pool, even if they are not active at that
moment), by using the script build\_machine\_file.sh in any machine in the 
HTCondor pool; and then move that file to the
same directory where you decompressed the files in step 1.



\item Install crontab entry to get the periodic data. That can be done in any
machine running HTCondor, but since you have to ensure it will be always
working, we recommend you install it on the HTCondor master node. If this
machine is different from the web server, make sure that the web server have
read access to the generate files.

\item Set the system up by setting the configuration variables with the 
appropriate values.

\end{enumerate}

\phantomsection
\section*{Acknowledgments} 

\addcontentsline{toc}{section}{Acknowledgments} 

We thank our colleagues in the IT Department who helped us with many small problems
related to HTCondor, especially issues with firewalls, HTCondor startup files, etc.
A big thank is due to {\'A}ngel de Vicente, who was the first to install and
maintain HTCondor in our Institute, and the main responsible for the big popularity
it is having, in term of usage, among our researchers.

\phantomsection
\bibliographystyle{unsrt}
\bibliography{sample}


\end{document}